\documentclass[9pt]{article}
\usepackage[utf8]{inputenc}
\usepackage{amsmath,amsfonts,amssymb,latexsym}
\usepackage[T1]{fontenc}
\newtheorem{satz}{Theorem}[section]

\newtheorem{defi}[satz]{Definition}

\newtheorem{bem}[satz]{Remark}

\newtheorem{koro}[satz]{Corollary}
\newtheorem{conclusion}[satz]{Conclusion}
\newtheorem{ob}[satz]{Observation}

\newcommand{\tit}{\textit}

\newcommand{\N}{\mathbb{N}}
\newcommand{\R}{\mathbb{R}}
\newcommand{\Z}{\mathbb{Z}}

\newcommand{\beq}{\begin{equation}}
\newcommand{\eeq}{\end{equation}}

\begin{document}
\thispagestyle{empty}
\begin{center}
\vspace*{1.0cm}
{\Large{\bf Subspaces of $\Z^n$ or $\R^n$ having Dimension $(n-\varepsilon)$ in the $(n-\varepsilon)$-Expansion}}
\vskip 1.5cm
{\large{\bf Manfred Requardt}}

\vskip 0.5cm

Institut fuer Theoretische Physik\\
Universitaet Goettingen\\
Friedrich-Hund-Platz 1\\
37077 Goettingen \quad Germany\\
(E-mail: requardt@theorie.physik.uni-goettingen.de) 

\end{center}
\begin{abstract} In the following we construct spaces of dimension $(n\pm \varepsilon)$ lying in the neighborhood of $\Z^n,\R^n$ in the context of the $(n-\varepsilon)$-expansion. We provide means and criteria to deform the spaces of integer dimension into this neighborhood. We argue that the field theoretic models living on these deformed spaces are the continuation of the models defined on the corresponding integer valued spaces. Furthermore
we perform the continuum limit of subgraphs of $\Z^n$ having non-integer dimension to the corresponding (fractal) subspaces of $\R^n$. We make sense of a fractal volume measure like $d^{(n-\varepsilon)}x$. 
\end{abstract}
\newpage
\section{Introduction} In the renormalization group framework an important role is played by the $\varepsilon$-expansion (see e.g. \cite{1},\cite{2},\cite{3}), that is, the formal extension from the integer dimension 4 to the noninteger dimension $(4-\varepsilon)$ and the representation of the various expressions as power series in $\varepsilon$. More generally, the general framework shows that, typically, we have a stable \tit{Gaussian fixed point}
for dimension $d>4$ which for $d<4$ becomes unstable while a \tit{non-Gaussian fixed point}, which was unstable for $d>4$, becomes the stable one for $d<4$.

The idea of the $\varepsilon$-expansion is to study what happens for small $\varepsilon=(4-d)$ and to derive power series for the \tit{critical exponents} in powers of $\varepsilon$. This has been done in a purely formal way without a deeper physical interpretation which would imply, among other things, the development of appropriate spaces of noninteger dimension. As Parisi expressed it in \cite{2}, either noninteger dimensional spaces are just a useful formal trick or they are a natural extension of integer dimensional spaces. In \cite{1} it was even remarked that the $\varepsilon$-expansion is a useful theoretical device without physical significance! 
 
Before we embark on an answer to this question some groundwork has to be done. For one, the role of dimension has to be analyzed in models of e.g. statistical physics, that is, the way it enters in the relevant expressions and, in a next step, the concept has to be appropriately generalized in order to cover a sufficiently large class of discrete and/or irregular spaces of integer and noninteger dimension. Of particular interest is the vicinity of spaces of integer dimension, that is, models defined on regular lattices like $\Z^n, \R^n,n\in\N$. More specifically, we want to develop a kind of \tit{deformation theory} of such regular spaces into nearby irregular spaces of dimension $n\pm\varepsilon$ which carry still some of the properties of for example $\Z^n$ or $\R^n$ as e.g. a certain homogeneity in the large.

There have been some atempts in the past to introduce ``lattices of effectively nonintegral dimensionality'' (\cite{Dhar}, see also \cite{Stillinger}) but the first systematic investigation on a broader scale, to our knowledge, appeared in \cite{Requ1}, proving, among other things, a number of characteristic properties of the dimension concept we introduced in \cite{Requ1} (at the time of writing \cite{Requ1} we were unaware of the paper \cite{Filk} in which the same notion was already used in a however more restricted way).

In \cite{Requ1} we argued that the important physical characteristic of a notion of dimension is the number of new! sites, which is reached after, say, n steps on a lattice compared to the number of sites reached after $(n-1)$ steps, starting from a fixed reference site. We then proceeded to extend this idea to arbitrary graphs or networks (see the following section) and proved a number of remarkable stability concepts of this notion under change of the starting point or perturbations of the graph geometry. We furthermore introduced a variety of different classes of graphs having almost arbitrary (non)integer dimensions.

We already mentioned in that paper the possibility of applying our results to the $\varepsilon$-expansion (a point of view also hold in \cite{Dhar}) but the typical examples of noninteger (fractal) dimension given in \cite{Requ1} or \cite{Dhar} were of a very inhomogeneous type, rather resembling the wellknown examples of fractal sets. As in our view dimension is only one of several characteristics of irregular spaces, we need more properties as e.g. a certain homogeneity in the large so that spaces can qualify as deformations of e.g. some $\Z^n$. That is, it cannot be expected that an arbitrary fractal space of noninteger dimension near some integer n is necessarily the appropriate stage for model systems lying in the vicinity of the corresponding model defined on $\Z^n$.

To better appreciate this crucial point it needs a deeper understanding and more technical tools to construct irregular spaces which really qualify as small deformations of some $\Z^n$. These tools were provided in more recent papers as e.g. \cite{Requ2} and \cite{Requ3}. A considerable stumbling block in the following analysis is a theorem we formulated in the above papers, i.e., the stability of dimension against many types of geometric deformations of the graph geometry. Put differently, if we want to deform e.g. $\Z^n$ into an irregular space of dimension $n\pm\varepsilon$ and which lies in a well-defined sense in the ``neighborhood'' of $\Z^n$, particularly subtle measures have to be taken. On the other hand, our hope is it  that this construction will lead to a better understanding of the whole subject matter and its subtleties.   
\section{Some Notions and Definitions}
Here are some notions and definitions from graph theory (we only introduce the absolute minimum; for more details see \cite{Requ1},\cite{Requ2} and the references given there). 
\begin{defi} A countable, labelled (unoriented) graph $G=(V,E)$ consists of a countable set of vertices (or nodes), $x_i\in V$, and a countable set of edges, $e_{ij}=(x_i,x_j)\in E\subset V\times V$, so that $e_{ij}$ and $e_{ji}$ are identified.
\end{defi}
\begin{defi} A vertex, x, has vertex degree $v(x)\in\N_0$ if it is incident with $v(x)$ edges. Such a vertex degree function is called locally bounded (in principle an infinite vertex degree is allowed).
\end{defi}
\begin{defi} A graph is called connected if each pair of vertices $x,y$ can be connected by a finite edge sequence, $\gamma$, starting at x and ending at y. An edge sequence (or walk) without repetion of vertices is called a path. The number of edges occurring in the path is called its length, $l(\gamma)$.
\end{defi}  
\begin{defi} Two graphs, $G,G'$, are called isomorphic if there exists a bijective map $\phi$ from $V$ to $V'$ which preserves adjacency.
\end{defi}
\begin{ob} As $l(\gamma)$ is an integer, there always exists a path of minimal length, which defines a (path) metric on G, i.e.:
\beq d(x,y):=\min_{\gamma}\{l(\gamma),\gamma\;\text{connects}\,x\,\text{with}\,y\}    \eeq
\end{ob} 

With the help of this metric we can define neighborhoods of vertices and various notions of dimension.
\begin{defi} The ball, B(x,r), of radius r around the vertex x is given by the vertices 
\beq y\in B(x,r)\;\text{if}\;d(x,y)\leq r\in\N_0     \eeq
The set of vertices with $d(x,y)=r$ is denoted by $\partial B(x,r)$. $|B(x,r)|$ and $|\partial B(x,r)|$ are the number of vertices lying in these sets.
\end{defi} 
\begin{defi} The growth function $\beta (G,x,r)$ is defined by 
\beq  \beta (G,x,r)=|B(x,r)|     \eeq
Correspondingly we define 
\beq \partial\beta (G,x,r)=\beta (G,x,r)-\beta (G,x,r-1)    \eeq
\end{defi}
\begin{defi}The (upper,lower) internal scaling dimension with respect
  to the vertex $x$ is given by
\begin{equation}\overline{D}_s(x):=\limsup_{r\to\infty}(\ln\beta(x,r) 
/\ln r)\;,\,\underline{D}_s(x):=\liminf_{r\to\infty}(\ln\beta(x,r)/\ln r)
\end{equation}
The (upper,lower) connectivity dimension is defined correspondingly as
\begin{equation}\overline{D}_c(x):=\limsup_{r\to\infty}(\ln\partial\beta(x,r)
/\ln r)+1\;,\,\underline{D}_c(x):=\liminf_{r\to\infty}(\ln\beta(x,r)/\ln
r)+1
\end{equation}
If upper and lower limit coincide, we call it the internal scaling
dimension, the connectivity dimension, respectively.
\end{defi}
Note that these accumulation points do always exist (with the value $\infty$ being included). While in most cases both notions coincide there exist examples where this is not the case (see \cite{Requ1}). For regular lattices these notions of dimension are the usual ones. Such behavior corresponds to the case of the many versions of fractal dimension.
\begin{bem} Sometimes such dimensions are called Hausdorff dimension (see e.g. \cite{Dur} or the discussion in the introduction of \cite{Requ3}). In our view such a designation is  problematical as it is actually quite the opposite of a fractal or Hausdorff dimension. While there exist some formal similarities the latter concepts typically describe the infinitely small structure of sets. On the other hand, our graph dimension describes the large scale behavior of the graph geometry. The two concepts becomes closer related when we perform a continuum limit of the graph geometry as described in e.g. \cite{Requ2} and \cite{Requ3}.
\end{bem}
To keep matters simple we treat in the following only the situation where all the  notions, introduced in the above definition are the same and where we denote this unique value by D. A simple corollary is the following:
\begin{koro}  For graphs with locally bounded vertex degree the above dimensional notions are independent of the vertex x.
\end{koro}

A nice and large class of graphs are those having \tit{polynomial growth}. 
\begin{defi} G is called to be of polynomial growth d if there are constants $A_x,B_x$ so that
\beq  A_xr^d\leq \beta(x,r)\leq B_xr^d     \eeq
with d the smallest possible value. One easily sees that in that case the dimension has the value d for all the above definitions.
\end{defi}

Extreme examples of graphs are regular trees (of vertex degree v) and regular lattices like e.g. $\Z^n$. In the first case we have
\beq \beta(r)=1+\sum_{\nu=0}^{r-1}\,(v-1)^{\nu}\sim v^r   \eeq
i.e., the dimension is $\infty$. For e.g. $\Z^2$ we have
\beq \beta(r)=2r^2+2r+1\leq Ar^2   \eeq
for some A. 
\begin{bem} Note that the graph metric for $\Z^n$ is not the euclidean one but the so-called $l^1$ or \tit{taxi cab} metric
\beq d_{l^1}(x,0)=\sum_{i=1}^n\,|x_i|\quad\text{for}\quad x\in\Z^n   \eeq
It is important to understand why the dimension is so large in the first case and so small in the latter case. Obviously it is not the vertex degree which matters but the number of different paths, starting e.g. at some initial vertex $0$ and terminating at some x, or, put differently, the number of closed paths, i.e. the connectivity of the graph! 
\end{bem}

In this paper we have to deal mainly with certain subgraphs of, say, $\Z^n$ having the following property:
\begin{defi} A subgraph $G'$ of a graph G is a spanning subgraph if it is defined on the same vertex set with its edge set being a subset of the edge set of G. Note that its induced metric $d_{G'}$ (defined again via the infimum over the length of paths in G') is in general different for given pairs $(x,y)$, i.e.
\beq  d_{G'}(x,y)\geq d_G(x,y)    \eeq

\end{defi}
\section{The Stability of Dimension under Deformations of the Graph Geometry}
We have seen that for example the vertex degree has almost no influence on the dimension of a graph. The same holds for other local deformations of the graph geometry as we will see immediately. This is the consequence of an important theorem, we proved in various versions in our above cited papers (see e.g. section 2 of \cite{Requ2} or more generally in \cite{Requ3}). We assume that we have an initial graph $G$ and begin to insert new additional edges or delete edges (while the vertex set remains fixed). It is easiest to assume that the edges are added or deleted all at once. We thus make a transition to a new graph $G'$. One would like to know how the dimension will change under such a deformation. 

The following theorem provides some insight.
\begin{satz} Insertions of arbitrarily many edges within an r-neighborhood of the vertices of G (with r a fixed value) do not change the dimension D of G. The same holds for edge deletions between vertices which have a distance smaller than some fixed r in the final graph $G'$. We call such operations r-local.
\end{satz}
\begin{bem} Note that the deletion process is exactly the inverse of the insertipn process, which can be seen if we start from the final $G'$ and reinsert the deleted edges. We tacitly assume of course that the graph $G'$ is still connected! Note furthermore that we always start from a graph G with locally bounded vertex degree. This implies that the maximal number of possible insertions within an r-neighborhood is always finite.
\end{bem}

These kinds of graph deformations can be considerably generalized to so-called \tit{quasi-isometries}. They played an important role in \cite{Requ2} and in particular in \cite{Requ3}.
\begin{defi}Let $F$ be a map from a metric space $X$  to a metric
  space $Y$ with metrics $d_X,d_Y$. It is called a quasi-isometric
  embedding if the following holds: There exist constants,
  $\lambda\geq 1,\epsilon\geq 0$, such that
\begin{equation}\lambda^{-1}\cdot d_X(x,y)-\epsilon\leq
  d_Y(F(x),F(y))\leq \lambda\cdot d_X(x,y)+\epsilon \end{equation}
If, furthermore, there exists a constant $\epsilon'$ such that for all $y\in
Y$ we have $d_Y(y,F(X))\leq \epsilon'$, that is, $Y\subset
U_{\epsilon'}(F(X))$ (the $\varepsilon'$-neighborhood of $F(X)$)
 it is called a quasi-isometry; the
spaces are then called quasi-isometric. There is an equivalent
definition which shows that the preceding definition is in fact
symmetric between $X$ and $Y$ (see for example \cite{Harpe}). That is,
there exists a quasi-isometric map $G$ from $Y$ to $X$ with
corresponding constants and $d_X(G\circ F(x),x)\leq\rho$ and
$d_Y(F\circ G(y),y)\leq\rho$ for some $\rho$. If $\lambda=1$ it is
called a rough isometry.
\end{defi}
We see that quasi-isometries allow us to compare metric spaces neglecting their small-scale structure and just looking at their coarse geometry.
\begin{koro} The above introduced edge insertions and deletions are quasi-isometries (cf. section 2 of \cite{Requ2}). 
\end{koro}
The concept of quasi-isometry plays a central role in \cite{Requ3} in the construction of a geometric renormalization group.

While quasi-isometries are more general than the above introduced graph deformations they still cannot change the dimension of a graph (if the following assumption is satisfied).
\begin{satz} If $G,G'$ are quasi-isometric graphs with globally bounded vertex degree, they have the same versions of dimension we defined above (this is theorem 2.22 in \cite{Requ2}; note however that it is theorem 2.16 in the arXiv-version).
 \end{satz}
We are now ready to attack the problem of constructing and describing the character of the spaces, having graph dimension $n\pm\varepsilon$, and living in the neigborhood of $\Z^n,\R^n$. 
\section{The $(n\pm\varepsilon)$- Neighborhood of $\Z^n$}
We start with some $\Z^n$ and want to add or delete edges in order to deform $\Z^n$ into some new discrete space. If we invoke a certain algorithm for deleting or inserting edges, we will most certainly arrive at a new graph, G, which has the same dimension as $\Z^n$ while it may nevertheless look quite  differently. The reason is that we typically are inclined to choose algorithms which are local in the sense defined in the preceding section, but we learned that such uniformly local changes do not alter the graph dimension. Only processes which are non-local of a certain type can change the dimension and these turn out to be quite difficult to describe.

We mentioned and discussed the (in our view) necessity of a certain \tit{translocal} structure in the fine structure of space-time in previous papers in models of quantum gravity or, rather, quantum space-time physics (see for example our remarks at the end of \cite{Requ3}, the \tit{critical network states} in \cite{Requ4} or the \tit{wormhole structure}  in \cite{Requ5}). It is helpful to have a pictorial model at our disposal which exhibits which kind of deletion/insertion process is necessary. In \cite{Requ1} or \cite{Requ4} we introduced the so-called \tit{Ulam-spiral} which consists of an ingeneous embedding of $\Z^1$ into $\Z^2$ in form of a spiral so that the vertex sets are the same and the vertices of $\Z^2$  can be labelled either by the coordinates of $\Z^2$ or $\Z^1$ . In other words, $\Z^1$ is mapped onto the spanning subgraph $U(\Z^1)\subset\Z^2$ with $U$ being an isometry of $\Z^1$ onto $U(Z^1)$, the latter with its induced distance metric.

 This graphical representation shows that in order to deform $\Z^2$ into $\Z^1$ or vice versa edges have to be deleted or inserted between vertices which have an increasingly large distance with respect to the labelling along the spiral, viz. $\Z^1$. We will employ this construction in the following analysis.
\begin{bem} The Ulam spiral plays a certain role in the context of the distribution of prime numbers.
\end{bem}
We begin our analysis by describing the needed nonlocal algorithms in an abstract way. To acomplish this we use our definitions of r-local deformations and quasi-isometry  and the theorem given in the previous section concerning the dimension of quasi-isometric graphs.

As the deformed graph G is a \tit{spanning subgraph} of $\Z^n$ in case of edge deletions  and is also given on the same vertex set in the case of edge insertions, we have a natural map $f$ (i.e. the identy map) from the vertex set of $\Z^n$ onto the vertex set of G. The above theorem states that this map $f:\Z^n\to G$ cannot be a quasi-isometry between the two metric spaces $\Z^n,G$ if their dimensions happen to be different. Furthermore, we learned that in that case the edge deletions/insertions cannot be r-local for any finite r. We begin with the discussion of edge deletions (we tacitly assume that the graph G remains connected).
\begin{ob}[edge deletions] As the map cannot be a quasi-isometry there always exist pairs of vertices $(x,x')$ for all $\lambda\in\R^+$ so that 
\beq  d_G(x,x') >\lambda\cdot d_{\Z^n}(x,x')    \eeq
(the other side of the estimate in the theorem is only relevant for edge insertions). This obviously implies that for any given fixed $\lambda$ there do exist infinitely many such pairs. Otherwise there would exist a finite $\lambda$ in the theorem.\\
Furthermore, as the edge deletions are not r-local for some r, for any given r there exist neighboring pairs $(x,x')$ in $\Z^n$ so that the edge between x and x' is deleted  but in $G$ we have 
\beq d_G(x,x')>r   \eeq
Again it holds that for any given fixed r there exist infinitely many such pairs.
\end{ob}
For edge insertions it correspondingly holds:
\begin{ob} [edge insertions] For all fixed $\lambda\in\R^+$ there exist infinitely many pairs $(x,x')$ so that 
\beq \lambda^{-1}\cdot d_{\Z^n}(x,x')\geq d_G(x,x')\quad\text{i.e.}\quad d_{\Z^n}(x,x')\geq\lambda\cdot d_G(x,x')    \eeq
and edges are inserted between vertices $(x,x')$ with
\beq d_{\Z^n}(x,x')>r   \eeq
for any $r>0$.
\end{ob}

We argued above that it is not sufficient to simply construct some graphs having non-integer dimensions lying between $n$ and $n\pm 1$. What we actually need are deformations of e.g. $\Z^n$ or $\R^n$ into their respective neighborhoods which are both ``controlled'' and ``continuous'' with respect to the parameter $\varepsilon$ in $n\pm\varepsilon$. Therefore we emphasize the following important point. In contrast to arbitrary possible constructions we restrict ourselves to the following scenario.
\begin{ob} We start from some $\Z^n$. The vertex sets remain fixed under the deformation. In the case of edge deletions we delete edges which exist in $\Z^n$. However, as we want to arrive at subgraphs which lie between $\Z^{n-1}$ and $\Z^n$ it is easier to start from $\Z^{n-1}$ and insert edges which belong to $\Z^n$.  In the case of edge insertions we insert only edges which exist in $\Z^{n+1}$.
\end{ob}  
We will see that this guarantees that our new spaces are really lying somehow between $\Z^n$ and $\Z^{n+1}$ or $\Z^{n-1}$. Note that in particular in the case of edge insertions this prescription strongly restricts our freedom of inserting new edges. Take e.g. the case of $\Z^1$. In general we could add new edges between arbitrary vertices but most of them would be incompatible with the postulate that the new graph is a spanning subgraph of $\Z^2$. This we will show below. In this way we may cover the whole intermediate space beween the end points $\Z^n,\Z^{n\pm 1}$. 

To perform this deformation process in a controlled and continuous way we will, in a first step, isometrically embed (with respect to the induced graph metric of the image graphs) $\Z^n$ in $\Z^{n+1}$ or $\Z^{n-1}$ in $\Z^n$ so that they all are given on the same vertex set as spanning subgraphs (as in the example of the Ulam spiral). In this way the vertices are either labelled by the coordinates of the spaces $\Z^n$ or $\Z^{n+1}$ or the embedded spaces $\Phi(\Z^{n-1}),\Phi(\Z^n)$ with $\Phi$ the embedding map, that is, the coordinates of $\Z^{n-1}$ or $\Z^n$. More specifically, let $(x_1,x_2,\ldots,x_n)$ be the coordinates of the vertices with respect to $\Z^n$ we write
\beq \Z^n=\Z^2\times\Z^{n-2}\;\text{with}\; (x_1,x_2)\in \Z^2, (x_3,\ldots,x_n)\in\Z^{n-2}    \eeq
We embed now $\Z^{n-1}$ in $\Z^n$ with the help of the Ulam embedding map, i.e.
\beq \Phi(\Z^{n-1})=U(\Z^1)\times\Z^{n-2}   \eeq
and correspondingly for $\Phi:\Z^n\to\Z^{n+1}$
\beq  \Phi(\Z^n)=U(\Z^1)\times\Z^{n-1}  \eeq

By inserting now further edges in $U(\Z^1)\subset \Z^2$ we can construct graphs lying between $\Z^n,\Z^{n+1}$ or $\Z^{n-1},\Z^n$, having dimensions $n\pm\varepsilon$.   
\begin{bem} The isometric map U, mapping $\Z^1$ onto a spanning subgraph of $\Z^2$, is of course not unique. But one can easily see that all these maps share the property that the image is strongly meandering through the embedding space $\Z^2$.
\end{bem}

The above embedding of e.g. $\Z^{n-1}$ in $\Z^n$ is a very special one. In general we envisage the following situation. 
\begin{ob} The general case of an isometric embedding is a bijective map $\phi$ from $\Z^{n-1}$ onto a spanning subgraph $\phi(\Z^{n-1})$ of $\Z^n$ which is isometric  with  $\phi(\Z^{n-1})$ carrying the induced graph metric.
\end{ob}
If we have a spanning subgraph $G$ in $\Z^n$ and want to argue that it lies between $\Z^{n-1}$ and $\Z^n$ we assume the following:
\begin{defi} We say that $G$ lies between $\Z^{n-1}$ and $\Z^n$ if the isometric embedding $\phi(\Z^{n-1})$ of $\Z^{n-1}$ in $\Z^n$ is contained as a spanning subgraph in $G\subset\Z^n$
\end{defi} 
\section{Spanning Subgraphs between $\Z^1$ and $\Z^2$}
In the following we want to analyze how insertions of additional edges in $U(\Z^1)$ lead to subgraphs having dimension $1<D\leq 2$. To this end we will describe some particular properties of $U(\Z^1)$ and the way it is embedded in $\Z^2$. 

The Ulam spiral starts at the point $0=(0,0)$ with the coordinate on the lhs belonging to $\Z^1$, the coordinate on the rhs to $\Z^2$. The next points or vertices are $1=(1,0)$ and $-1=(-1,0)$ followed by $2=(1,1), -2= (-1,-1), 3=(0,1),-3=(0,-1)$. In this way the Ulam spiral winds around the point $(0,0)$ counterclockwise with alternating turns consisting of positive or negative $\Z^1$-coordinates. We will label these turns by counting their intersection with the vertical positive axis $(0,0)-(0,\infty)$. We denote the horizontal coordinate by $x$ and the vertical coordinate by $y$. That is, a general coordinate in $\Z^2$ is denoted by $(x,y)$.

This alternation of turns with positive and negative $\Z^1$-coordinates has the effect that while the vertical $\Z^2$-distance of nearest neighbors (nn) on $U(\Z^1)$ is one, their distance along the Ulam spiral (i.e. their $\Z^1$-distance) strongly increases with every turn. To illustrate this important point we give some numbers  for the vertical $y$-line $(0,0)-(0,\infty)$. In the vertical direction we have the $\Z^1$-coordinates:
\beq  0,3,-10,21,-36,54  \eeq
for the $\Z^2$-labelled vertices:
\beq (0,0),(0,1),(0,2),(0,3),(0,4)   \eeq
That is, the $\Z^1$-distance increases like
\beq  3,13,31,57,90   \eeq 
\begin{ob} This is exactly the behavior we decribed in the general theorems of section 3. As the Ulam spiral (or $\Z^1$) is given on the same vertex set as $\Z^2$ but has dimension 1 instead of 2, it cannot be quasi-isometric to $\Z^2$. That is, there exist nn's in $\Z^2$ which have arbitrarily large distance with respect to $U(\Z^1)$ viz. $\Z^1$.
\end{ob}

We now start our construction of subgraphs of $\Z^2$. Our first example is  the spanning subgraph constructed from $U(\Z^1)$ and the line $(0,0)-(0,\infty)$, that is we add the edges of the line $x=0$ to the edges of $U(\Z^1)$ and denote this new graph by $G$. While this graph has still far lesser edges compared to $\Z^2$, we will show that its dimension is nevertheless already 2. We proved in e.g. \cite{Requ1} that the dimension of a locally finite graph is independent of the reference vertex. Thus, for convenience we choose $(0,0)$ as reference vertex and undertake to calculate $\partial\beta((0,0),r)$ as defined in section 3.
\begin{bem} It turns out that in our case the calculation of the graph dimension D is much simpler by using $\partial\beta((0,0),r)$ instead of  $\beta((0,0),r)$ which we will show below.
\end{bem}
That is, we use the definition
\beq D:=\lim_{r\to\infty}(\ln\partial\beta((0,0),r)/\ln(r))+1     \eeq
with
\beq \partial\beta((0,0),r)=\beta((0,0),r)-\beta((0,0),r-1)   \eeq
being the number of vertices which can be reached in $r$ steps but not in $(r-1)$ steps.
\begin{bem} We showed in \cite{Requ1} that the two definitions of $D$ are not always the same but coincide in the ordinary cases like ours.
\end{bem}

To calculate or estimate  $\partial\beta((0,0),r)$ we begin by making $r$ steps along the line $x=0$, reaching the vertex $(0,r)$. In order to facilitate the argumentation we discuss the case $(0,r)$. The case $(0,-r)$ gives, due to symmetry, only a factor 2 in the final formula. This line segment $(0,0)-(0,r)$ (or $(0,0)-(0,-r)$, respectively) crosses the Ulam spiral $r$ times. The behavior of $\partial\beta$ in each step from $(r-1)$ to $r$ is quite transparent. For $r=1$ we get two new vertices on the Ulam spiral proper and the vertex $(0,1)$. For distance $r$ at each vertex $(0,r')$ with $r'<r$ we get two new vertices by moving a  step to the left on the respective turn of the Ulam spiral and a step to the right. Thus we have made $r-r'$ steps to each side.
\begin{bem} Note the following slight complication. For $r$ sufficiently large we have made $(r-r)'$ steps on the $r'$-th turn of the Ulam spiral and $(r-r'-1)$ on the $(r'-1)$ -th turn. For small $r'$ it will thus happen that after sufficiently many steps all the vertices between $(0,r')$ and $(0,r'-1)$ are finally reached and we do not get more new vertices on this part of the Ulam spiral.
\end{bem}
We will compensate this numerical complication in the following way.

The formula for the dimension of a graph is very stable against small perturbations due to the occurring logarithms and the taking of the limit $r\to\infty$. It is hence sufficient to take only sufficiently large distances $r$ into account. We have shown above that the distance between consecutive vertices $(0,r)$ and $(0,r+1)$ along the Ulam spiral is strongly growing. That is, we can restrict our analysis on values $r/2<r'\leq r$ or choosing an appropriate constant $0<C<1$ instead of $1/2$. We then have  
\begin{ob} For sufficiently large $r$ or $r\to\infty$ it holds
\beq \partial\beta((0,0),r)\gtrsim 4C\cdot r\quad\text{for some}\quad 0<C<1  \eeq
This implies that $D$ is  $\leq 2$ but $\geq \lim_{r\to\infty}(\ln(4C\cdot r)/\ln(r))+1=2$
\end{ob}
\begin{conclusion} The spanning subgraph $G$ has the same dimension as $\Z^2$ itself.
\end{conclusion}

We see that this example of a spanning subgraph of $\Z^2$ does not lead to a dimension being smaller than 2. Furthermore we note the following:
\begin{ob} This example shows that two graphs can have the same dimension while not being quasi-isometric.
\end{ob}
On the other hand this example enables us to understand what is needed to construct subgraphs with dimension $D=2-\varepsilon$. In the above example we added a straight line to the Ulam spiral (in fact any (tilted) line will do). That is, we cross a new turn of $U(\Z^1)$ after every step $(r\to r+1)$. Thus the number of new vertices increases proportional to $r$. This is exactly the problem.

We can avoid this by choosing instead of a straight line a sufficiently meandering curve $\gamma(r)$ which we parametrize by its length $r$, i.e., the number of edges. We make the following choices. The curve starts again at $(0,0)$ and goes to infinity, i.e.
\beq  |\gamma(r)|:=|x(r)|+|y(r)|\to\infty\quad\text{with}\quad r\to\infty  \eeq
Furthermore the curve is assumed to be a path, that is, a vertex on the curve is never met twice. The crucial property is the following: The length of the curve, $r$, fulfills
\beq r\gg |\gamma(r)|\quad\text{for sufficiently large}\quad r  \eeq
This implies that $\gamma(r)$ does no longer simply cross the Ulam spiral but a certain fraction of the edges on  $\gamma(r)$ are lying on $U(\Z^1)$.

In principle we can choose arbitrary meandering curves but we want to simplify the discussion a little bit by assuming that $\gamma(r)$ is lying in the forward cone given by
\beq |x|-|y|=0\;,\;y\geq 0  \eeq
This can be assumed because we are at the moment only interested in small deviations $\varepsilon$ in $D=2-\varepsilon$. We make the further assumption that it holds $y(r+1)\geq (y(r)$, i.e. we have either that $y(r+1)=y(r)$ or $y(r+1)=y(r)+1$. It follows:
\begin{ob} $y(r)$ is the number of turns of the Ulam spira, $\gamma(r)$ has crossed. 
\end{ob}

We now choose a curve with $y(r)=r^{1-\varepsilon}$, that is, $r^{\varepsilon}$ edges of $\gamma(r)$ are lying on $U(\Z^1)$. For sufficiently small $\varepsilon$ all the arguments we have given in the preceding example remain valid and we have
\begin{conclusion} For such a subgraph of $\Z^2$ it holds
\beq D=\lim_{r\to\infty} \ln(C\cdot r^{1-\varepsilon})/\ln(r)+1=2-\varepsilon  \eeq
\end{conclusion}
\section{The Continuum Limit of Spanning Subgraphs of $\Z^n$ with Dimension $(n-\varepsilon)$}
In quantum field theory and other fields of modern physics it is interesting to have spaces of dimension  $(n-\varepsilon)$ at ones disposal, that is, spaces which are contained in $\R^n$. The spaces which we constructed above are spanning subgraphs of $\Z^n$. By construction they have \tit{polynomial growth} $(n-\varepsilon)$ as defined in section 2. We proved in \cite {Requ2} and \cite{Requ3} with the help of a detailed analysis that such spaces have a continuum limit in the sense of Gromov-Hausdorff (GH). This continuum limit was generated by employing the sequence of graph metrics 
\beq d_l(x,y):=l^{-1}\cdot d(x,y)  \eeq
with $d$ the standard graph metric and $l\to\infty$. One can envisage these spaces as subgraphs of the lattices $l^{-1}\Z^n$, i.e., with the shrinking lattice spacing  $l^{-1}\cdot 1$.

As $\Z^n$ has $\R^n$ as GH-limit the GH-limit of these subgraphs $G\subset\Z^n$ can be understood as spaces being embedded in $\R^n$ (as sets) but presumably in a complicated \tit{metrical} way. In general they can be expected to be of a \tit{fractal type}. Some of their properties have been proved in \cite{Requ2} and \cite{Requ3}. However, in general, their concrete shape is difficult to envisage. By the same token it is not easy to develop some form of general analysis on such spaces. For certain examples of fractal spaces this is done in e.g. \cite{Stri}. In the following we will employ the special strucuture of the spaces under discussion as limits of subspaces of some $\Z^n$ or $l^{-1}\Z^n$. 

In continuum field theory one is frequently interested in propagators and correlation functions and the integration of such expressions over certain subsets of $\R^n$. In the $\varepsilon$-expansion the integration with respect to the standard Riemann or Lebesgue measure, i.e. $d^nx$, is replaced in a purely formal way by something like $d^{n-\varepsilon}x$. More specifically, this means that certain integral expressions are calculated with the help of the ordinary Lebesgue measure $d^nx$ and then in the final expression the occurring dimensional parameter $n$ is formally replaced by $(n-\varepsilon)$ wherever it occurs. We now want to justify this procedure by analyzing the limits of our subspaces of $l^{-1}\Z^n$ with $l$ large or $l\to\infty$.  

Up to now the examples of such spaces, i.e., spanning subgraphs of some $\Z^n$, are not very isotropic in all directions in a, of course, coarse sense. That is, we selected some sublattice $\Z^2$, constructed an embedded subgraph $\phi(\Z^1)\subset\Z^2$ of dimension $(2-\varepsilon)$ with the help of the Ulam spiral and then formed the cartesian product $\Z^{n-2}\times\phi(\Z^1)\subset\Z^n$.

If we construct the continuum limit of spaces having fractal dimension $(n-\varepsilon)$, we would of course prefer on physical grounds to arrive at limit spaces lying in $\R^n$ which are homogeneous and isotropic in some approximate (i.e. large scale) sense. This implies that we should start from subgraphs $G$ of $\Z^n$ which are already homogeneous and isotropic in a coarse sense. At the moment it appears to be too difficult to explicitly construct such types of subgraphs $G$, having dimension $(n-\varepsilon)$ and contain $\Z^{n-1}$ as embedded subgraph, so that one can say that $G$ lies between $\Z^{n-1}$ and $\Z^n$.  

So, for the rest of this section, we will assume that such subgraphs $G\subset\Z^n$ do exist. Due to the general results we mentioned above, $G$ has a $GH$-limit lying in $\R^n$ but has a quite complicated metrical fine structure. Our idea in the following is to circumvent, for the time being, the problem of describing this fractal fine structure by choosing a more qualitative approach, with the help of which we remain within the realm of $\R^n$. We want to describe some properties of the fractal fine structure of the limit spaces by projecting these properties into the ordinary euclidean space $\R^n$. 

To this end we start from $l^{-1}\Z^n$ and the embedded subspace $l^{-1}G$ with $G$ the subgraph having dimension $(n-\varepsilon)$. The integration in $\R^n$ we describe with the help of polar cordinates with infinitesimal solid angle $d\Omega$ and infinitesimal volume element $d^nx=r^{(n-1)}drd\Omega$. On $\R^n$ we want to use the euclidean distance. Note that on $\Z^n$ the canonical graph distance metric is the $l^1$- or taxi cab metric
\beq d_{\Z^n}(x,0)=\sum_{i=1}^n|x_i|  \eeq

We now choose a spherical shell in $\R^n$  lying between $r$ and $r+\Delta r$ with $\Delta r$ small compared to $r$. 
\beq \Delta B^n_r:=B^n_{r+\Delta r}\setminus B^n_r   \eeq
With $l$ sufficiently large the points of $l^{-1}\Z^n$ lying in $\Delta B^n_r$ are densely distributed with density $\rho^l_{\Z^n}= l^{-n}$ on $\R^n$. For the number of points $N(r,\Omega)$ of $l^{-1}\Z^n$ lying in $r^{(n-1)}\Delta r\Delta\Omega$ it holds:
\beq   N_l(r,\Omega)\approx r^{(n-1)}\Delta r\Delta\Omega/l^{-n}  \eeq
It is more convenient in the following to employ the euclidean metric $d_E$ also on $\Z^n$. It is easy to see that $d_E$ and $d_G$ differ on $\Z^n$ only by a small numerical factor.

We now come to the subgraph $l^{-1}G\subset  l^{-1}\Z^n$. According to our general observations made in the preceding sections we know that there has to exist a substantial fraction of pairs of vertices $(x,y)$ with
\beq d_G(x,y)\gg d_{\Z^n}(x,y)   \eeq
in order that the dimension of $G$ is smaller than $n$. The above relation happens to be scaled in   $l^{-1}G\subset  l^{-1}\Z^n$ but, on the other hand, for large $l$ we can move to vertices having larger and larger distance in $G$ so that in the limit $l\to\infty$ we arrive at points having arbitrary finite  distance and being densely distributed in $\R^n$.

$G$ has the property that the number of vertices having distance $d_G(x,0)=r$ increases like
\beq \partial\beta (G,0,r)\sim r^{(n-\varepsilon)-1}  \eeq
The number of points of $l^{-1}\Z^n$ lying in $r^{(n-1)}\Delta r\Delta\Omega$ for large $l$ is given by formula (33). As in $G$ edges have been deleted we have
\beq d_E^G(0,x)\geq d_E^{\Z^n}(0,x)  \eeq
We assumed that the vertices, $x$, in $G$, having distance $d_E^G(0,x)=r$ from the reference vertex $0$, are evenly distributed and have growth degree $\sim r^{(n-1)-\varepsilon}$. They all lie in $B_{r+\Delta r}\subset\R^n$.
\begin{bem} Note again that in our discussion we have to carefully distinguish between vertices in $G$ having graph distance $r$ from $0$ and their euclidean or $l^1$-distance as points embedded in $\R^n$.

Furthermore, we regard $l^{-1}G\subset l^{-1}\Z^n$ as (lattice) graphs embedded in $\R^n$ with edge length $l^{-1}$. That is, the euclidean graph distance $d_l^{\Z^n}$ on $l^{-1}\Z^n$ is the ordinary euclidean distance in $\R^n$ which is, however, not! the case for the subgraph $G$.
\end{bem}

We now make the transition $l\to\infty$ and $\Delta r\to dr,\Delta\Omega\to d\Omega$ with $l$ chosen so large that there still lie sufficiently many points of $l^{-1}\Z^n$ in
\beq r^{(n-1)}\Delta r\Delta\Omega\to r^{(n-1)}drd\Omega   \eeq 
We collect the points of $l^{-1}G$ lying within the solid angle $\Delta\Omega$ and having euclidean graph distance $d_E^G$ from $0$ between $r$ and $r+\Delta r$ and distribute them evenly in the small volume element $ r^{(n-1)}\Delta r\Delta\Omega$ (note again that  $d_E^G$ is different from the euclidean distance on $\R^n$).

By construction this density is less than $\rho_l^{\Z^n}=l^{-n}$. We have instead that 
\beq \rho_l^G=r^{-\varepsilon}l^{-n}  \eeq
Taking now the limit $l\to\infty$ and $ r^{(n-1)}\Delta r\Delta\Omega\to r^{(n-1)}drd\Omega$ we get:
\begin{ob} The limit space $\lim_{l\to\infty} l^{-1}G$ can be characterized in $\R^n$ by an infinitesimal volume element
\beq  r^{(n-1)-\varepsilon}drd\Omega   \eeq
instead of the ordinary euclidean volume element $r^{(n-1)}drd\Omega$.
\end{ob}
This new volume element can be used instead of the ordinary one in the integration of the respective expressions occurring in e.g. field theories on these embedded fractal limit spaces.

\end{document}